\documentclass[numbered]{trbunofficial}
\usepackage{graphicx}
\usepackage{xcolor,colortbl}
\usepackage{caption}
\usepackage{subcaption}
\usepackage[hidelinks]{hyperref}

\definecolor{Gray}{gray}{0.3}

\AuthorHeaders{Wang, Wei, Lin, and Li}
\title{Spatial-temporal Analysis of COVID-19's Impact on Human Mobility: the Case of the United States}

\author{%
  \textbf{Songhe Wang$^*$}\\
  University of North Carolina at Chapel Hill\\
  songhe17@live.unc.edu\\
  \hfill\break
  \textbf{Kangda Wei$^*$}\\
  University of North Carolina at Chapel Hill\\
  kangda@live.unc.edu\\
  \hfill\break%
  \textbf{Lei Lin}\\
  University of Rochester\\
  lei.lin@rochester.edu\\
  \hfill\break%
  \textbf{Weizi Li}\\
  University of Memphis\\
  wli@memphis.edu
}

\begin{document}
\maketitle

\section{Abstract}
COVID-19 has been affecting every aspect of societal life including human mobility since December, 2019. In this paper, we study the impact of COVID-19 on human mobility patterns at the state level within the United States. From the temporal perspective, we find that the change of mobility patterns does not necessarily correlate with government policies and guidelines, but is more related to people's awareness of the pandemic, which is reflected by the search data from Google Trends. Our results show that it takes on average 14 days for the mobility patterns to adjust to the new situation. From the spatial perspective, we conduct a state-level network analysis and clustering using the mobility data from Multiscale Dynamic Human Mobility Flow Dataset. As a result, we find that 1) states in the same cluster have shorter geographical distances; 2) a 14-day delay again is found between the time when the largest number of clusters appears and the peak of Coronavirus-related search queries on Google Trends; and 3) a major reduction in other network flow properties, namely degree, closeness, and betweenness, of all states from the week of March 2 to the week of April 6 (the week of the largest number of clusters).

\hfill\break%
\noindent\textit{Keywords}: COVID-19, Human Mobility, Spatial-temporal Analysis, Network Analysis, Clustering
\newpage

\section{Introduction}

The COVID-19 pandemic has been affecting the world since December, 2019~\cite{huang2020clinical}. As for the United States, the government announced a national emergency on March 13 and attempted to slow down the spread of the virus by issuing social distancing guidelines. Many societal aspects are subsequently affected including human mobility---the driven force for social-economic development. According to Apple COVID-19 Mobility Trends Reports~\cite{Apple}, after declaring national emergency, the traffic flow, including driving and walking, has decreased over 50\%. 


The impact of COVID-19 on the mobility of various social-demographic groups within the United States has been assessed by several studies~\cite{kabiri2020different,ghader2020observed,engle2020staying,lee2020human,fellows2020covid,hu2020impacts,sun2020covid,galeazzi2020human,dahlberg2020effects,askitas2020lockdown}. To provide some examples, Kabiri et al.~\cite{kabiri2020different} study the difference of mobility changes of younger and senior communities. They find that both younger and senior communities have been performing social distancing after the national emergency declaration. The percentage of people staying at home and the social distancing index, a measurement based on the change of a community’s daily work trips, show a trend of decreasing until mid-April and then gradual increasing. In addition, they find that people need a certain period of time to react to government guidelines and term this phenomenon \emph{social distancing inertia}~\cite{ghader2020observed}. Engle et al.~\cite{engle2020staying} evaluate how local infection rate would lead to mobility reduction. They show that the region with a higher percentage of residents older than 65 has a higher mobility reduction rate. Lee et al.~\cite{lee2020human} analyze the relationship between mobility and various socio-demographic factors and find that 1) the community with higher income has a higher stay-at-home percentage compared to the community with lower income after the outbreak, and 2) the community with higher population density has shorter travel distance compared to the community with lower population density. Fellows et al.~\cite{fellows2020covid} conclude that even though many factors, such as perceived risk, community guidance, and media coverage, could affect people’s mobility patterns, the factor that has the most effect is the governmental non-pharmaceutical intervention, e.g., the staying-at-home order. Similar studies have also been conducted in other countries in analyzing the impact of COVID-19 on human mobility patterns~\cite{galeazzi2020human,dahlberg2020effects,askitas2020lockdown}. 

Five months after the declaration of national emergency, many states have moved to various phases of reopening. Some studies have examined this process. For example, Huang et al.~\cite{huang2020population} study the impact of reopening on mortality, infections, and hospitalizations in Harris county, Texas, United States. As another example, Soltani and Rezazadeh~\cite{soltani2020new} suggest that releasing a small portion of the population may cause a significant increase in confirmed cases. Most previous work focused on different responses from various social-demographic communities and the local factors associated with the mobility patterns. Others study the effect of reopening on the pandemic's evolution. However, these studies do not study the relationship between people's awareness of the situation and the change of mobility patterns, and do not explore the mobility patterns between different states. Our work not only explores the correlation between people's awareness of the pandemic and the mobility patterns, but also uses network analysis and clustering to reveal the underlying mobility structures between different states. 

To be specific, from the temporal aspect, we analyze the correlation between human mobility and the amount of search queries from Google Trends at the state level. As a result, we find that, on average, people adjust their mobility two weeks after their awareness of the situation. From the spatial aspect, we perform network clustering to reveal hidden mobility structures among all states. The result shows that the connections among states at first decreased and then gradually restored after mid-April. Meanwhile, the states in the same cluster tend to have shorter geographically distances. In addition, we find, again, a two-week delay between the point when the largest number of clusters appears and the point when the peak of the Google search trend regarding Coronavirus occurs, which confirms our temporal analysis results. Lastly, the network analysis shows a reduction of degree, closeness, and betweenness---the properties that describe mobility flows of a state---for all states.

In the rest of the paper, we first discuss the data sources used in our analysis. Then, we explain our methodology for this research. Next, we present temporal and spatial analysis results and explain our findings. Finally, we conclude and discuss potential future work.

\section{Datasets}

In this section, we provide details of our data sources. 

Regarding the mobility data, the first dataset that we use is Apple COVID-19 Mobility Trends Report~\cite{Apple}. The dataset contains traffic flows measured by the number of route changing queries via Apple Map of all states of the U.S., from January 13 to July 20, 2020. The flow of January 13 is set to 100, and the flow of other days is computed relatively to the flow data of January 13. For example, if the traffic flow of a certain day is increased by 20\% compared to the traffic flow of January 13, the flow of that day is marked as 120. Some sample data are shown in Table~\ref{table:apple-data}.

\begin{table}[h!]
\centering
\scalebox{.85}{
 \begin{tabular}{cccccccccccc} 
 \hline
 State & 1/13/20 & 1/14/20 & 1/15/20 & ... & 3/14/20 & 3/15/20 & 3/16/20 & ... & 7/18/20 & 7/19/20 & 7/20/20 \\ 
 \hline\hline
 Alabama & 100 & 102.9 & 103.51 & ... & 139.87 & 94.6 & 101.37 & ... & 204.26 & 142.92 & 151.62\\ 
 Arizona & 100 & 104.06 & 106.9 & ... & 113.76 & 85.38 & 94.73 & ... & 112.83 & 91.34 & 101.85\\
 Arkansas & 100 & 102.55 & 101.93 & ... & 113.61 & 83.41 & 90.28 & ... & 181.21 & 138.92 & 151.08 \\
 California & 100 & 104.39 & 109.34 & ... & 86.16 & 66.24 & 77.57 & ... & 119.08 & 101.23 & 107.6\\
 Colorado & 100 & 103.46 & 105.66 & ... & 98.41 & 75.88 & 81.94 & ... & 160.1 & 138.23 & 143.85\\
 \hline
 \end{tabular}}
 \caption{Samples of daily traffic data of the U.S. from Apple COVID-19 Mobility Trends Report~\cite{Apple}. The flow of January 13 is set to 100 and the flows of other days are computed relatively to the flow of January 13. The national emergency is declared on March 13 in the U.S.}
 \label{table:apple-data}
\end{table}

The second dataset is from Google Trends~\cite{Google}, which we use as an indicator of people's awareness of the pandemic. The dataset contains both state-level and nationwide weekly search data regarding Coronavirus. We extract the part of the data from January 13 to July 20, 2020 for our analysis. The data are adjusted: with 0 meaning a very small number of queries was made and 100 representing the largest number of queries made over the past year. A number between 0 and 100 indicates that a week has a specific higher search percentile than all other weeks over the past year. Some sample data are shown in Table~\ref{table:google-data}. Furthermore, in order to conduct daily analysis, we linearly interpolate the weekly data so that each day in the study period also has the search data metric.

\begin{table}[h!]
\centering
\scalebox{.8}{
 \begin{tabular}{cccccccccccc} 
 \hline
 Region & 1/12/20 & 1/19/20 & 1/26/20 & ... & 3/8/20 & 3/15/20 & 3/22/20 & ... & 7/12/2020 & 7/19/2020 & 7/26/2020 \\ 
 \hline\hline
Alabama & $<1$ & 3 & 8 & ... & 67 & 100 & 81 & ... & 10 & 7 & 7\\ 
Arizona & $<1$ & 4 & 19 & ... & 62 & 100 & 83 & ... & 14 & 11 & 9\\ 
Arkansas & $<1$ & 4 & 13 & ... & 74 & 100 & 79 & ... & 11 & 8 & 7\\ 
California & $<1$ & 5 & 21 & ... & 74 & 100 & 75 & ... & 13 & 10 & 9\\ 
Colorado & $<1$ & 5 & 12 & ... & 87 & 100 & 75 & ... & 11 & 9 & 8\\ 
 \hline\hline
Nationwide & <1 & 4 & 13 & ... & 73 & 100 & 74 & ... & 11 & 9 & 8 \\ 
 \hline
 \end{tabular}}
 \caption{Samples of weekly search data from Google Trends regarding Coronavirus~\cite{Google}. 0 meaning the lowest search amount over the past year and 100 meaning the highest search amount over the past year.  As a reference,  the national emergency is declared on March 13 in the U.S.}
 \label{table:google-data}
\end{table}


The third dataset is the Multiscale Dynamic Human Mobility Flow Dataset~\cite{kang2020multiscale}, which contains weekly mobility flow data, provided by SafeGraph~\cite{SafeGraph}, from the week of March 2 to the week of May 11, 2020. Each week's data contain the origin and destination of visitor and population flow, and the longitudes and latitudes of the origin states and destination states. The visitor flow is measured by recording the geographical locations of mobile devices, and the population flow is calculated based on the visitor flow and the population of each state. 

Regarding the pandemic data, we use the daily reports of COVID-19 state-level confirmed cases from New York Times~\cite{nytimes}. We have pre-processed all the data by conducting min-max feature scaling. This operation ensures that the data are on the same scale, providing convenience for our modeling, analysis, and visualization.

\section{Spatial-temporal Analysis}

\subsection{Mobility Adjustment}


We hypothesize that there exists a time difference between the change of mobility and people's awareness of the pandemic---which is assumed to be reflected by Google Trends data. In other words, we suspect that the point of people realizing the seriousness of a situation and the point of people taking actions to reflect their awareness may not be well synchronized. 

\begin{figure}[ht]
  \centering
  \includegraphics[width=.8\textwidth]{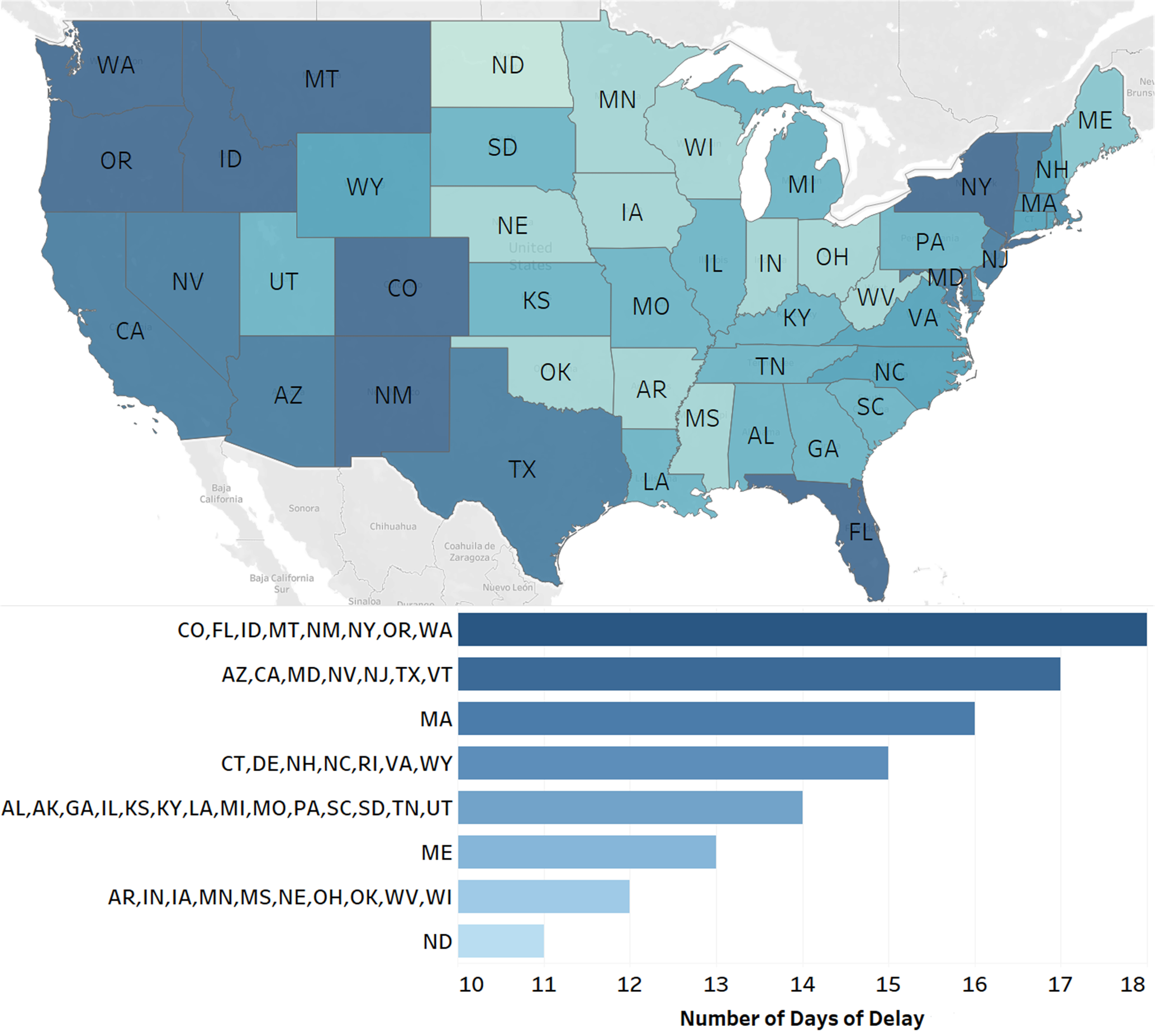}
  \caption{The number of days of delay reflected by Apple mobility data~\cite{Apple} to Google search data~\cite{Google} for all states. The average delay of people's mobility adjustment to their awareness of the pandemic is 14 days.}
  \label{fig:delay}
\end{figure}

To test our hypothesis, we study the correlation between the two time series: the mobility data from Apple~\cite{Apple} and the search trend data of COVID-19 from Google~\cite{Google}. To be specific, we compute the Pearson's correlation coefficients by fixing the mobility data and shifting the search trend data one day at a time. The study period is from January 13 to July 20, 2020. As a result, the highest correlation coefficients (absolute values) all occur after shifting the search trend data forward in time. The number of days of delay corresponds to the highest correlation coefficients of different states is shown in Figure~\ref{fig:delay}. The minimum delay of people adjusting their mobility to reflect their awareness of the pandemic is 11 days, the maximum delay is 18 days, and the average delay is 14 days. These observations coincide with the previous finding from Soltani and Rezazadeh~\cite{soltani2020new}, which states that people need a certain amount of time to adjust their mobility to the posed government guidelines.   

\subsection{Network Analysis and Clustering}
In order to explore possible hidden structures embedded in the mobility flow network among all states~\cite{lin2014data}, we perform network clustering using the Label Propagation algorithm implemented in the Python library NetworkX~\cite{hagberg2008exploring} on the Multiscale Dynamic Human Mobility Flow Dataset~\cite{kang2020multiscale}. The graph we study contains 50 states as its nodes and the traffic flows among the states as its edges. The algorithm works as follows: first the algorithm assigns each node a distinct label and then recursively assigns the label of each node to be the label that appears most frequently among the labels of the node's neighbors. In the case of multiple labels sharing the highest frequency, one of the labels will be randomly chosen for the node. The algorithm stops when all nodes have the labels that appear most frequently among the labels of their neighbors.

Since the Multiscale Dynamic Human Mobility Flow Dataset contains weekly inter-state mobility flows over 11 weeks (from the week of March 2 to the week of May 11, 2020), we have 11 mobility flow networks to analyze. For each week, we run the Label Propagation algorithm 100 times to reveal hidden clusters. During that process, we record the mode, mean, and standard deviation of the number of the clusters of the 11 weeks. These statistics are shown in Table~\ref{table:cluster}. We take the mode of the number of clusters of each week as the final number of clusters to conduct further analysis. 

\begin{table}[h!]
\centering
\scalebox{0.75}{
 \begin{tabular}{cccccccccccc} 
 \hline
 Statistics & Week 1 & Week 2 & Week 3 & Week 4 & Week 5 & Week 6 & Week 7 & Week 8 & Week 9 & Week 10 & Week 11 \\ 
 \hline\hline
 Mode & 1 & 1 & 2 & 5 & 6 & 8 & 7 & 6 & 5 & 5 & 4 \\ 
 Mean & 1 & 1 & 2.44 & 4.85 & 6.49 & 7.66 & 7.34 & 6.66 & 5.45 & 4.36 & 3.96 \\ 
 Std & 0 & 0 & 1.12 & 0.89 & 1.28 & 1.11 & 1.25 & 1.21 & 1.14 & 0.99 & 1.19 \\ 
 \hline
 \end{tabular}}
 \caption{Statistics of the clustering results using weekly mobility flow data from the Multiscale Dynamic Human Mobility Flow Dataset~\cite{kang2020multiscale}. The larger the number of clusters, the more isolated the states are. The largest number of clusters appears at week 6, i.e., the week of April 6.}
 \label{table:cluster}
\end{table}

\begin{figure}[ht]
  \centering
  \includegraphics[width=\textwidth]{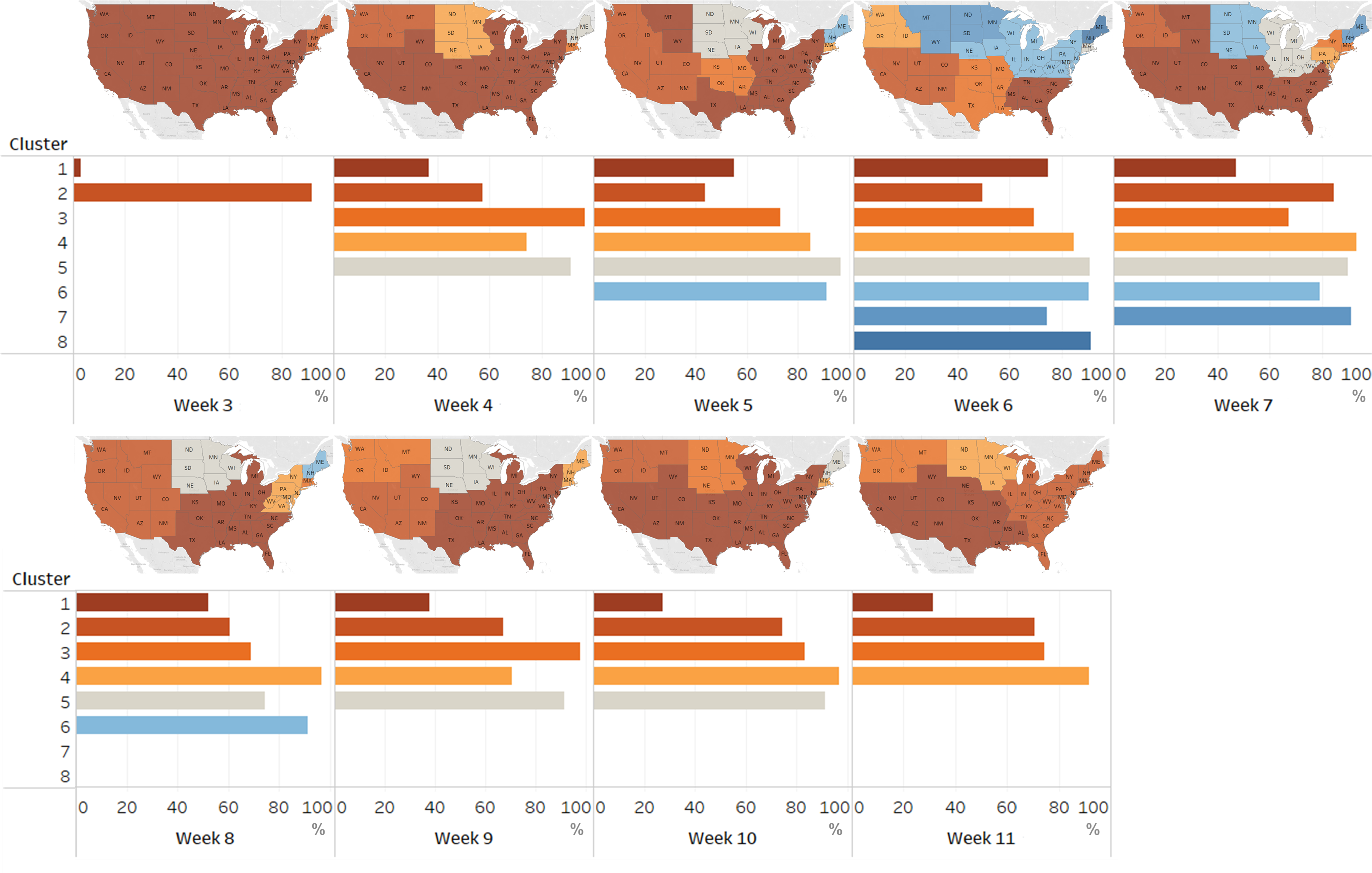}
  \caption{Clustering results starting from the week of March 16 and ending at the week of May 11. The visualization results, as well as the relative reduction in percentage of cluster average distance compared to the country average distance are shown. The cluster average distances are much shorter than the country average distances, indicating the states in the same cluster are closer to each other, which may be the results of shortened travel distances of people due to the pandemic. }
  \label{fig:cluster}
\end{figure}

From the clustering result, we observe that the number of clusters tend to increase and then decrease over the study period of 11 weeks. To be specific, in the first two weeks (before the national emergency declaration on March 13), all states are identified as one cluster. We suspect the reason being that people's awareness of COVID-19 remained low and the staying-at-home order was not implemented. As the pandemic evolves and people become more aware of the situation, the number of clusters increases, which implies people have restricted their travel and the states are more isolated. The maximum number of clusters appeared at week 6 (the week of April 6). This result may correlate with some major events occurred between March 13 and April 6, including the declaration of national emergency, the issuing of stay-at-home orders in all 50 states, and the cancellation of events with crowds. Afterwards, the number of clusters starts to decrease, which implies that people gradually resume their prior-pandemic travel behaviors. Figure~\ref{fig:cluster} TOP showcases the clustering results. To provide some examples, Week 3 (the week of March 16) shows that in the beginning of the pandemic, we only have two clusters, which indicates that mobility flows are more uniformly distributed across the country. Week 6 (the week of April 6) presents the largest number of clusters, i.e., eight clusters, indicating that the mobility flows among states reach the lowest point. Week 11 (the week of May 11) shows that the number of clusters reduces to four, showing that the states are less isolated and the mobility flows start to recover.

In order to quantify our results, we further compute two types of distances: \emph{cluster average distance} and \emph{country average distance}. Both distances are calculated using the coordinates of the geometric centroids of the states. For computing the cluster average distance, we first empirically pick a state in the cluster as the centroid and then take the average of the distances from the rest of the states in that cluster to the centroid. Next, we compute the country average distance by taking the distances from all states to the centroid. The relative reduction in percentage of cluster average distance compared to the country average distance of all clusters over nine weeks, starting from the week of March 16 and ending at the week of May 11, are shown in Figure~\ref{fig:cluster}. The first two weeks, i.e., the weeks of March 2 and March 9 are exclude since for these two weeks the algorithm reports the whole country as one cluster. As a result, all cluster average distances are much shorter than their corresponding country average distances, which suggests that the states in the same cluster are geographically close to each other. This result confirms that the increase of the number of clusters indeed reflects the isolation of the states and reductions in inter-state mobility flows. 

Table~\ref{table:cluster-and-search} shows the change of the number of clusters and nationwide weekly Google Trends data over the 11 weeks. A two-week delay between the peak of search trends on Google and the largest number of clusters is again spotted. It is worth noticing that even though we use different mobility data in our mobility adjustment analysis and network analysis, both analyses find the same amount of delay in time between the mobility change and people's awareness of the pandemic.

\begin{table}[h!]
\centering
\scalebox{0.72}{
 \begin{tabular}{cccccccccccc} 
 \hline
  & Week 1 & Week 2 & Week 3 & Week 4 & Week 5 & Week 6 & Week 7 & Week 8 & Week 9 & Week 10 & Week 11 \\ 
 \hline\hline
 Num. of Cluster(s) & 1 & 1 & 2 & 5 & 6 & \cellcolor{gray}8 & 7 & 6 & 5 & 5 & 4 \\ 
  \hline
 Google Trends & 16 & 31 & 73 & \cellcolor{gray}100 & 74 & 59 & 57 & 43 & 34 & 20 & 18 \\ 
 \hline
 \end{tabular}}
 \caption{The change of the number of clusters and weekly Google Trends data over the 11 weeks. We observe a two-week delay between the peak of Google Trends data and the largest number of clusters. This result coincides with our previous finding, which, on average, it takes 14 days for people to adjust their mobility to reflect their awareness of the pandemic (indicated by the Google Trends data). }
 \label{table:cluster-and-search}
\end{table}

Because the pandemic may affect different states unequally, we have examined other network flow properties, namely the degree, closeness, and betweenness of each state as a node in the traffic flow network. In particular, the degree of a node is defined as the sum of edge weights to all its neighbors. The change of the degree can inform the change of the flow of a state. The closeness of a node is defined as the average length of the shortest path (computed using the edge weights) from the node to all other nodes in the graph, which can reflect the accessibility of a state. Having the closeness calculated for all nodes, we could study which states tend to be less ``attractive'' during the pandemic. Lastly, betweenness quantifies the number of times a node that acts as the ``bridge'' along the shortest path between two other nodes. The betweenness of a node can be used to identify which state is serving as a domestic transportation ``hub''. The higher the value the more influence the state has on the domestic mobility.  

As part of our quantitative analysis, Table~\ref{table:top} lists top 10 states with most reductions in degree, closeness, and betweenness of the week of April 6 (the week has the largest number of clusters and most isolated) compared to the week of March 2 (the beginning of the pandemic). We can see that the average reductions in degree, closeness, and betweenness are 7.7\%, 6.9\%, and 21.6\% respectively. 

\begin{figure}[ht]
  \centering
  \includegraphics[width=\textwidth]{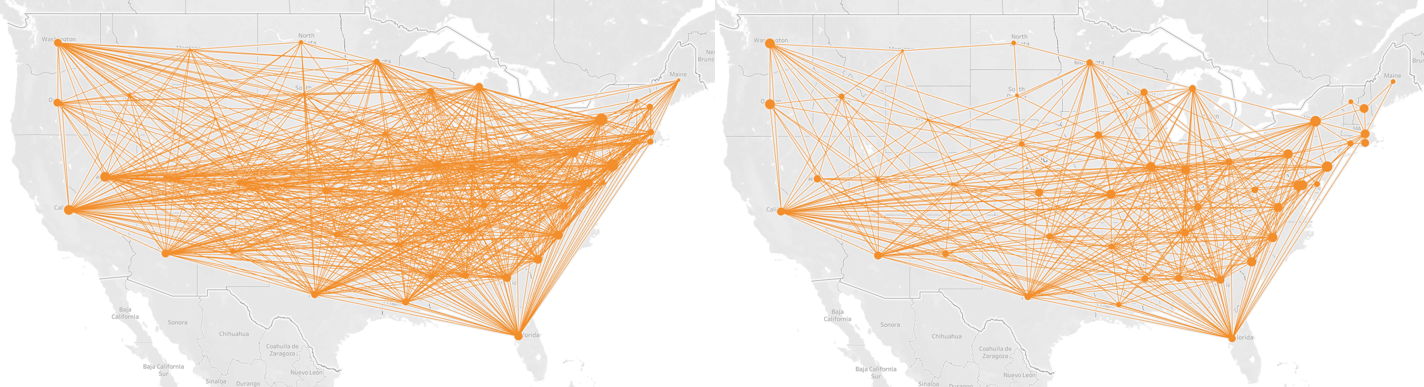}
  \caption{The inter-state flow connections of the week of March 2 (LEFT) and the week of April 6 (RIGHT). Each edge represents 10,000+ flow connecting two states. A significant reduction of the mobility flow is observed comparing the two weeks. }
  \label{fig:connection}
\end{figure}

As part of our qualitative analysis, Figure~\ref{fig:connection} visualizes the change of inter-state flow connections of the week of March 2 and the week of April 6. The size of a node represents the total mobility flow through that state during the week, while each edge indicates 10,000+ mobility flow connecting two states. A significant reduction of the edges (i.e., mobility flow) is observed.

\begin{table}[ht]
\centering
\scalebox{.72}{
 \begin{tabular}{cc|cc|cc} 
 \hline
 State & Degree Reduction & State & Closeness Reduction & State & Betweenness Reduction \\ 
 \hline\hline
 Nevada & 32.5\% & New Hampshire & 19.3\% & Nevada & 97.8\% \\ 
 Massachusetts & 25.5\% & Maine & 19.0\% & Wisconsin & 57.1\% \\ 
 District of Columbia & 24.3\% & Rhode Island & 15.4\% & New Jersey & 53.7\% \\
 New York & 22.7\% & Massachusetts & 14.8\% & North Dakota & 44.3\% \\ 
 Maine & 22.5\% & Nevada & 12.0\% & New Hampshire & 43.2\% \\ 
 New Hampshire & 21.6\% & California & 10.9\% & Colorado & 32.4\% \\ 
 Arizona & 20.9\% & Arizona & 10.7\% & Arizona & 30.1\% \\ 
 Colorado & 18.5\% & District of Columbia & 10.3\% & Utah & 29.9\% \\ 
 Wisconsin & 16.0\% & North Dakota & 9.6\% & Michigan & 22.3\% \\ 
 Vermont & 15.4\% & New Jersey & 9.1\% & Idaho & 20.4\% \\ 
 \hline
 Mean & 7.7\% & Mean & 6.9\% & Mean & 21.6\% \\
 Std & 9.7\% & Std & 4.0\% & Std & 27.3\% \\
 \hline
 \end{tabular}}
 \caption{Top 10 states with most reductions in degree, closeness, and betweenness of the week of April 6 (the most isolated week in terms of mobility flow) compared to the week of March 2.}
 \label{table:top}
\end{table}

\section{Conclusion}
The outbreak of COVID-19 has significantly impacted our society and dramatically changed our mobility patterns. We conduct spatial-temporal analysis of the influence of COVID-19 on human mobility in the United States. From the temporal perspective, we find that there exists a delay of, on average, 14 days between people's awareness of the pandemic and the change of mobility patterns. From the spatial perspective, we discover that the states within the same cluster tend to have shorter geographical distances. There also exists a two-week delay between the appearing of the largest number of clusters and the occurrence of the peak of the Google search trend regarding Coronavirus. Lastly, we perform a network analysis and find that there is a reduction of the network flow properties, namely degree, closeness, and betweenness, for all states.

For the future research, we are mainly interested in two directions. The first direction concerns a more detailed mobility analysis at the city level. We can first use existing techniques such as Compressed Sensing~\cite{Li2017CityEstSparse,Lin2019ComSense} to interpolate the sparse mobile data used in this work. In combination with other COVID-19 tracker data, we can then potentially estimate the impact of COVID-19 on city-level traffic states and simulate the dynamics of the virus spread using some existing techniques~\cite{Li2018CityEstIter,Wilkie2015Virtual,Li2017CityFlowRecon}. Another interesting topic in this direction is to study the change of shared mobility such as bikes~\cite{Lin2019BikeTRB,lin2018predicting}. The second research direction concerns future mobility, as the future traffic system is projected to be connected and autonomous~\cite{future}, we would like to study when mobility patterns have a drastic change, to what extent the connected and autonomous vehicles can be used to supplement the traffic system and fulfill various societal needs. We can leverage recent advances in autonomous driving modeling and simulation to facilitate this research direction~\cite{Li2019ADAPS,Chao2019Survey,Lin2021Attention}.

\bibliographystyle{trb}
\bibliography{references}

\begin{thebibliography}{31}
\providecommand{\natexlab}[1]{#1}

\bibitem[{Huang et~al.(2020{\natexlab{a}})Huang, Wang, Li, Ren, Zhao, Hu,
  Zhang, Fan, Xu, Gu et~al.}]{huang2020clinical}
Huang, C., Y.~Wang, X.~Li, L.~Ren, J.~Zhao, Y.~Hu, L.~Zhang, G.~Fan, J.~Xu,
  X.~Gu, et~al., Clinical features of patients infected with 2019 novel
  coronavirus in Wuhan, China. \emph{The lancet}, Vol. 395, No. 10223,
  2020{\natexlab{a}}, pp. 497--506.

\bibitem[{App(2020)}]{Apple}
\emph{{COVID-19}-Mobility Trends Reports}.
  \url{https://covid19.apple.com/mobility}, 2020, accessed: 2020-08-29.

\bibitem[{Kabiri et~al.(2020)Kabiri, Darzi, Zhou, Sun, and
  Zhang}]{kabiri2020different}
Kabiri, A., A.~Darzi, W.~Zhou, Q.~Sun, and L.~Zhang, How different age groups
  responded to the COVID-19 pandemic in terms of mobility behaviors: a case
  study of the United States. \emph{arXiv e-prints}, 2020.

\bibitem[{Ghader et~al.(2020)Ghader, Zhao, Lee, Zhou, Zhao, and
  Zhang}]{ghader2020observed}
Ghader, S., J.~Zhao, M.~Lee, W.~Zhou, G.~Zhao, and L.~Zhang, Observed mobility
  behavior data reveal social distancing inertia. \emph{arXiv preprint
  arXiv:2004.14748}, 2020.

\bibitem[{Engle et~al.(2020)Engle, Stromme, and Zhou}]{engle2020staying}
Engle, S., J.~Stromme, and A.~Zhou, Staying at home: mobility effects of
  COVID-19. \emph{Available at SSRN}, 2020.

\bibitem[{Lee et~al.(2020)Lee, Zhao, Sun, Pan, Zhou, Xiong, and
  Zhang}]{lee2020human}
Lee, M., J.~Zhao, Q.~Sun, Y.~Pan, W.~Zhou, C.~Xiong, and L.~Zhang, Human
  Mobility Trends during the COVID-19 Pandemic in the United States.
  \emph{arXiv preprint arXiv:2005.01215}, 2020.

\bibitem[{Fellows et~al.(2020)Fellows, Slayton, and Hakim}]{fellows2020covid}
Fellows, I.~E., R.~B. Slayton, and A.~J. Hakim, The COVID-19 Pandemic,
  Community Mobility and the Effectiveness of Non-pharmaceutical Interventions:
  The United States of America, February to May 2020. \emph{arXiv preprint
  arXiv:2007.12644}, 2020.

\bibitem[{Hu et~al.(2020)Hu, Barbour, Samaranayake, and Work}]{hu2020impacts}
Hu, Y., W.~Barbour, S.~Samaranayake, and D.~Work, Impacts of COVID-19 mode
  shift on road traffic. \emph{arXiv preprint arXiv:2005.01610}, 2020.

\bibitem[{Sun et~al.(2020)Sun, Zhou, Kabiri, Darzi, Hu, Younes, and
  Zhang}]{sun2020covid}
Sun, Q., W.~Zhou, A.~Kabiri, A.~Darzi, S.~Hu, H.~Younes, and L.~Zhang, COVID-19
  and Income Profile: How People in Different Income Groups Responded to
  Disease Outbreak, Case Study of the United States. \emph{arXiv preprint
  arXiv:2007.02160}, 2020.

\bibitem[{Galeazzi et~al.(2020)Galeazzi, Cinelli, Bonaccorsi, Pierri, Schmidt,
  Scala, Pammolli, and Quattrociocchi}]{galeazzi2020human}
Galeazzi, A., M.~Cinelli, G.~Bonaccorsi, F.~Pierri, A.~L. Schmidt, A.~Scala,
  F.~Pammolli, and W.~Quattrociocchi, Human Mobility in Response to COVID-19 in
  France, Italy and UK. \emph{arXiv preprint arXiv:2005.06341}, 2020.

\bibitem[{Dahlberg et~al.(2020)Dahlberg, Edin, Gr{\"o}nqvist, Lyhagen,
  {\"O}sth, Siretskiy, and Toger}]{dahlberg2020effects}
Dahlberg, M., P.-A. Edin, E.~Gr{\"o}nqvist, J.~Lyhagen, J.~{\"O}sth,
  A.~Siretskiy, and M.~Toger, Effects of the covid-19 pandemic on population
  mobility under mild policies: Causal evidence from sweden. \emph{arXiv
  preprint arXiv:2004.09087}, 2020.

\bibitem[{Askitas et~al.(2020)Askitas, Tatsiramos, and
  Verheyden}]{askitas2020lockdown}
Askitas, N., K.~Tatsiramos, and B.~Verheyden, Lockdown strategies, mobility
  patterns and COVID-19. \emph{arXiv preprint arXiv:2006.00531}, 2020.

\bibitem[{Huang et~al.(2020{\natexlab{b}})Huang, Chu, Shams, Kim, Allen,
  Annapragada, Subramanian, Kakadiaris, Gottlieb, and
  Jiang}]{huang2020population}
Huang, T., Y.~Chu, S.~Shams, Y.~Kim, G.~Allen, A.~V. Annapragada,
  D.~Subramanian, I.~Kakadiaris, A.~Gottlieb, and X.~Jiang, Population
  stratification enables modeling effects of reopening policies on mortality
  and hospitalization rates. \emph{arXiv preprint arXiv:2008.05909},
  2020{\natexlab{b}}.

\bibitem[{Soltani and Rezazadeh(2020)}]{soltani2020new}
Soltani, K. and G.~Rezazadeh, A New Dynamic Model to Predict the Effects of
  Governmental Decisions on the Progress of the CoViD-19 Epidemic. \emph{arXiv
  preprint arXiv:2008.11716}, 2020.

\bibitem[{Goo(2020)}]{Google}
\emph{Coronavirus Search Trends}.
  \url{https://trends.google.com/trends/story/US_cu_4Rjdh3ABAABMHM_en}, 2020,
  accessed: 2020-09-13.

\bibitem[{Kang et~al.(2020)Kang, Gao, Liang, Li, Rao, and
  Kruse}]{kang2020multiscale}
Kang, Y., S.~Gao, Y.~Liang, M.~Li, J.~Rao, and J.~Kruse, Multiscale Dynamic
  Human Mobility Flow Dataset in the US during the COVID-19 Epidemic.
  \emph{arXiv preprint arXiv:2008.12238}, 2020.

\bibitem[{Saf(2020)}]{SafeGraph}
\emph{{S}afe{G}raph}. \url{https://www.safegraph.com/}, 2020, accessed:
  2020-12-05.

\bibitem[{nyt(2020)}]{nytimes}
\emph{Coronavirus (Covid-19) Data in the United States}.
  \url{https://github.com/nytimes/covid-19-data}, 2020, accessed: 2020-09-14.

\bibitem[{Lin et~al.(2014)Lin, Wang, and Sadek}]{lin2014data}
Lin, L., Q.~Wang, and A.~W. Sadek, Data mining and complex network algorithms
  for traffic accident analysis. \emph{Transportation Research Record}, Vol.
  2460, No.~1, 2014, pp. 128--136.

\bibitem[{Hagberg et~al.(2008)Hagberg, Swart, and
  S~Chult}]{hagberg2008exploring}
Hagberg, A., P.~Swart, and D.~S~Chult, \emph{Exploring network structure,
  dynamics, and function using NetworkX}. Los Alamos National Lab.(LANL), Los
  Alamos, NM (United States), 2008.

\bibitem[{Li et~al.(2017{\natexlab{a}})Li, Nie, Wilkie, and
  Lin}]{Li2017CityEstSparse}
Li, W., D.~Nie, D.~Wilkie, and M.~C. Lin, Citywide Estimation of Traffic
  Dynamics Via Sparse {GPS} Traces. \emph{IEEE Intelligent Transportation
  Systems Magazine}, Vol.~9, No.~3, 2017{\natexlab{a}}, pp. 100--113.

\bibitem[{Lin et~al.(2019{\natexlab{a}})Lin, Li, and Peeta}]{Lin2019ComSense}
Lin, L., W.~Li, and S.~Peeta, Efficient Data Collection and Accurate Travel
  Time Estimation in a Connected Vehicle Environment via Real-Time Compressive
  Sensing. \emph{Journal of Big Data Analytics in Transportation}, Vol.~1,
  No.~2, 2019{\natexlab{a}}, pp. 95--107.

\bibitem[{Li et~al.(2018)Li, Jiang, Chen, and Lin}]{Li2018CityEstIter}
Li, W., M.~Jiang, Y.~Chen, and M.~C. Lin, Estimating urban traffic states using
  iterative refinement and Wardrop equilibria. \emph{IET Intelligent Transport
  Systems}, Vol.~12, No.~8, 2018, pp. 875--883.

\bibitem[{Wilkie et~al.(2015)Wilkie, Sewall, Li, and Lin}]{Wilkie2015Virtual}
Wilkie, D., J.~Sewall, W.~Li, and M.~C. Lin, Virtualized Traffic at
  Metropolitan Scales. \emph{Frontiers in Robotics and AI}, Vol.~2, 2015,
  p.~11.

\bibitem[{Li et~al.(2017{\natexlab{b}})Li, Wolinski, and
  Lin}]{Li2017CityFlowRecon}
Li, W., D.~Wolinski, and M.~C. Lin, City-Scale Traffic Animation Using
  Statistical Learning and Metamodel-Based Optimization. \emph{ACM Trans.
  Graph.}, Vol.~36, No.~6, 2017{\natexlab{b}}, pp. 200:1--200:12.

\bibitem[{Lin et~al.(2019{\natexlab{b}})Lin, Li, and Peeta}]{Lin2019BikeTRB}
Lin, L., W.~Li, and S.~Peeta, Predicting Station-Level Bike-Sharing Demands
  Using Graph Convolutional Neural Network. In \emph{Transportation Research
  Board 98th Annual Meeting (TRB)}, 2019{\natexlab{b}}.

\bibitem[{Lin et~al.(2018)Lin, He, and Peeta}]{lin2018predicting}
Lin, L., Z.~He, and S.~Peeta, Predicting station-level hourly demand in a
  large-scale bike-sharing network: A graph convolutional neural network
  approach. \emph{Transportation Research Part C: Emerging Technologies},
  Vol.~97, 2018, pp. 258--276.

\bibitem[{fut(2020)}]{future}
\emph{The car of the future is connected, autonomous, shared, and electric}.
  \url{https://www.zdnet.com/article/the-car-of-the-future-is-connected-autonomous-shared-and-electric/},
  2020, accessed: 2020-08-29.

\bibitem[{Li et~al.(2019)Li, Wolinski, and Lin}]{Li2019ADAPS}
Li, W., D.~Wolinski, and M.~C. Lin, {ADAPS}: Autonomous Driving Via Principled
  Simulations. In \emph{IEEE International Conference on Robotics and
  Automation (ICRA)}, 2019, pp. 7625--7631.

\bibitem[{Chao et~al.(2019)Chao, Bi, Li, Mao, Wang, Lin, and
  Deng}]{Chao2019Survey}
Chao, Q., H.~Bi, W.~Li, T.~Mao, Z.~Wang, M.~C. Lin, and Z.~Deng, A Survey on
  Visual Traffic Simulation: Models, Evaluations, and Applications in
  Autonomous Driving. \emph{Computer Graphics Forum}, Vol.~39, No.~1, 2019, pp.
  287--308.

\bibitem[{Lin et~al.(2021)Lin, Li, Bi, and Qin}]{Lin2021Attention}
Lin, L., W.~Li, H.~Bi, and L.~Qin, Vehicle Trajectory Prediction Using LSTMs
  with Spatial-Temporal Attention Mechanisms. \emph{IEEE Intelligent
  Transportation Systems Magazine}, 2021.

\end{thebibliography}
\end{document}